\documentclass[epj,final,amsmath,amssymb]{svjour}

\usepackage{graphicx}
\usepackage{amssymb}

\input{epsf}

\begin{document}

\title {Two-dimensional ranking of Wikipedia articles}

\author{ A.O.Zhirov \inst{1}, O.V.Zhirov \inst{2} and 
D.L.Shepelyansky \inst{3} }
\institute{Novosibirsk State University, 630090 Novosibirsk, Russia
\and
Budker Institute of Nuclear Physics, 
630090 Novosibirsk, Russia
\and
Laboratoire de Physique Th\'eorique du CNRS, IRSAMC, 
Universit\'e de Toulouse, UPS, F-31062 Toulouse, France
}

\titlerunning{Two-dimensional ranking of Wikipedia}
\authorrunning{A.O.Zhirov, O.V.Zhirov and D.L.Shepelyansky}

\abstract{The Library of Babel, described by   
Jorge Luis Borges, stores an enormous
amount of information. The Library exists {\it ab aeterno}. 
Wikipedia,  a free online encyclopaedia, becomes a modern 
analogue of such a Library.
Information retrieval and ranking of Wikipedia articles become
the challenge of modern society. 
While PageRank highlights very well known nodes with many ingoing links,
CheiRank highlights very communicative nodes with many outgoing links.
In this way the ranking becomes two-dimensional.
Using CheiRank and PageRank 
we analyze the properties of two-dimensional ranking
of all Wikipedia English articles and show that it gives their reliable 
classification with rich and nontrivial features. 
Detailed studies are done for
countries, universities, personalities, physicists, chess players, 
Dow-Jones companies and other categories.
}

\date{Received: June 28, 2010; Revised: September 20, 2010}

\maketitle

\section{Introduction}

The {\it Encyclop\'edie} \cite{encyclopedie} 
accumulates the available human
  knowledge making it accessible to all {\it citoyennes}. In this way the
{\it Encyclop\'edie} becomes one of the most powerful catalysts of modern
development of science and society \cite{blom}. This process of knowledge
transfer becomes enormously accelerated with the appearance of Wikipedia 
\cite{wikipedia}, a free online encyclopaedia, which current size overcomes
6 millions English entries \cite{wikipedia,wikimedia}.
Wikipedia comes close to 
Encyclopaedia Britanica \cite{britanica} in terms of
the accuracy of its science entries \cite{giles} overcoming the later 
by far in an enormous amount of available information. 
The classification and ranking of this information 
becomes the great challenge.
The statistical analysis of directed network
generated by links between Wikipedia articles \cite{zlatic,caldarelli,muchnik}
established their scale-free properties showing certain similarities 
with the World Wide Web and other scale-free networks
\cite{watts,newman,barabasi,dorogovtsev,meyer}. 
Here we apply a two-dimensional 
ranking algorithm, proposed recently \cite{alik}, and
classify all Wikipedia articles in English by their degree of communication
and popularity. This ranking allows to select 
articles in a new way, giving e.g. more preference 
to communicative and artistic 
sides of a personality compared to popularity and political aspects stressed
by the PageRank algorithm \cite{brin}. With a good reliability
the ranking of Wikipedia articles
reproduces the well established classifications
of countries \cite{country}, universities \cite{shanghai},
personalities \cite{hart}, 
physicists, chess players, Dow-Jones companies \cite{dowjones}
and other categories.

The paper is composed as follows: in Section 2 we describe
CheiRank and PageRank algorithms and apply them for two-dimensional 
ranking of Wikipedia articles, in Section 3 the results of such a ranking
are discussed for various categories of articles,
discussion of findings is given in Section 4.

\section{CheiRank versus PageRank}

\begin{figure}
\begin{center}
\includegraphics[clip=true,width=4.0cm]{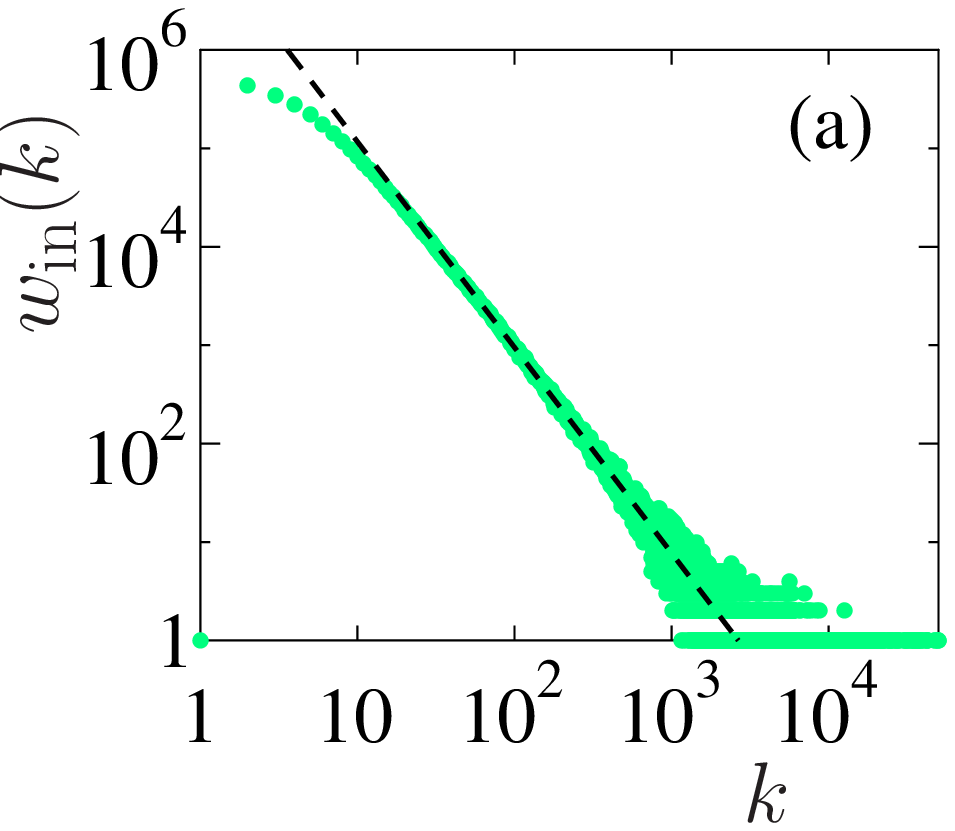}
\includegraphics[clip=true,width=4.0cm]{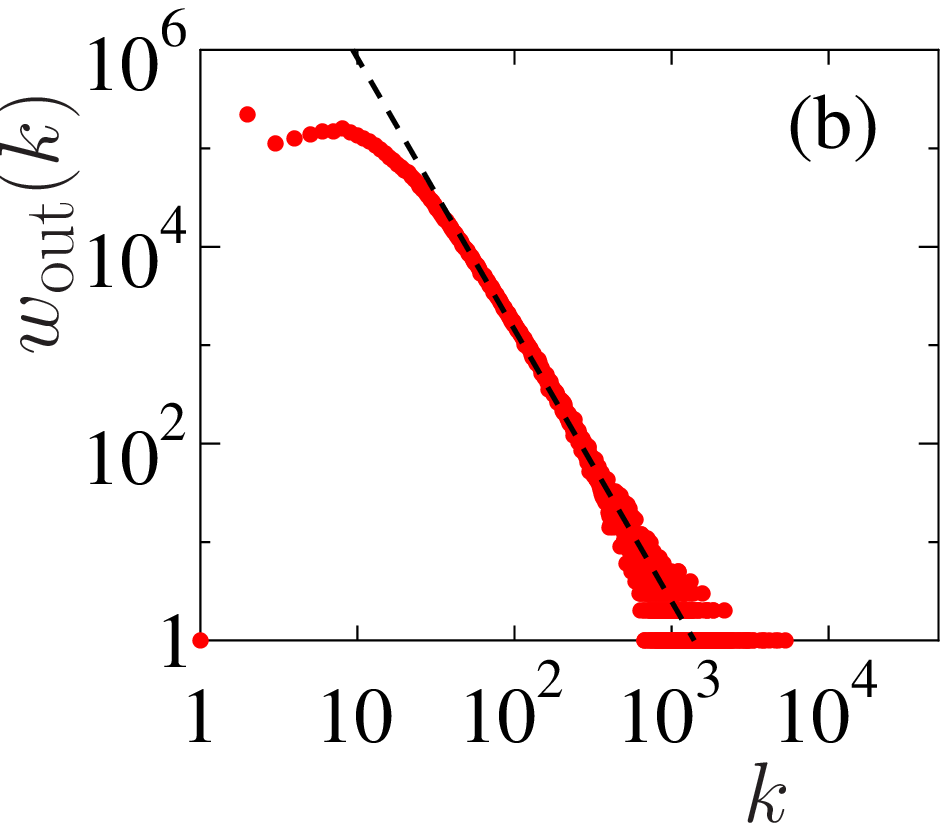}
\vglue -0.1cm
\caption{(Color online) 
Distribution $w_{in,out}(k)$ of ingoing (a)
and outgoing (b) links $k$ for $N=3282257$
Wikipedia English articles. The straight dashed fit line shows the slope
with $\mu_{in}=2.09 \pm 0.04$ (a) and 
$\mu_{out}=2.76 \pm 0.06$ (b).
}
\label{fig1}
\end{center}
\end{figure}

At August 18, 2009 we downloaded from \cite{wikimedia} 
the latest English Wikipedia snapshot and
by crawling determined links between all $N_{tot}=6855098$ entries.
We keep for our study only entries which are linked with other entries
and eliminate categories from consideration.  After that we obtain a directed
network of $N=3282257$ articles. The multiplicity of links from one article to
another is taken into account. The distributions of ingoing
and outgoing  links are well described by 
a power law $w_{in,out}(k) \propto 1/k^{\mu_{in,out}}$
with the exponents $\mu_{in}=2.09 \pm 0.04$ 
and $\mu_{out}=2.76 \pm 0.06$ 
(see Fig.~\ref{fig1})
being in a good agreement with the previous studies of Wikipedia
\cite{zlatic,caldarelli,muchnik} and the World Wide Web (WWW) 
\cite{meyer,donato,upfal}.

\begin{figure}
\begin{center}
\includegraphics[clip=true,width=8.2cm]{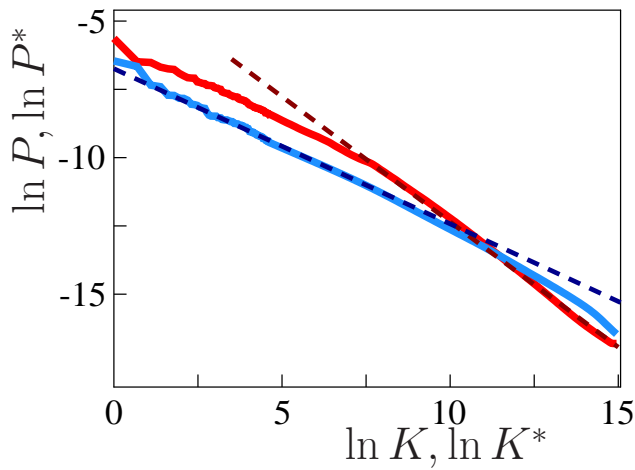}
\vglue -0.1cm
\caption{(Color online) 
Dependence of probability of PageRank $P$ (red curve) and
CheiRank $P^*$ (blue curve) 
on the corresponding rank indexes $K$ and $K^*$. 
The straight dashed lines show
the power law dependence with the slope
$\beta=0.92; 0.57$ respectively,  corresponding to 
$\beta=1/(\mu_{in,out}-1)$.
}
\label{fig2}
\end{center}
\end{figure}

Due to the similarity with the WWW it is natural to construct 
the Google matrix of Wikipedia using the procedure
described in \cite{brin,meyer}:
\begin{equation}
  G_{ij} = \alpha S_{ij} + (1-\alpha) / N \;\; ,
\label{eq1}
\end{equation} 
where the matrix $S$ is obtained by normalizing 
to unity all columns of the adjacency matrix,
and replacing columns with zero elements by $1/N$, $N$ 
being the network size.
The damping parameter $\alpha$ in the WWW context describes the probability 
to jump to any node for a random surfer. 
For Wikipedia this parameter can describe the 
probability to modify an article that affects the overall ranking. 
The value $\alpha = 0.85$  gives
a good classification \cite{meyer} for WWW
and  thus we also use this value here.
The matrix $G$ belongs to the class of Perron-Frobenius operators \cite{meyer},
its largest eigenvalue 
is $\lambda = 1$ and other eigenvalues have $|\lambda| \le \alpha$. 
The right eigenvector 
at $\lambda = 1$ gives the probability $P(i)$ to find 
a random surfer at site $i$ and
is called the PageRank. Once the PageRank is found, 
Wikipedia articles are sorted by decreasing $P(i)$, 
the article rank in this index $K(i)$
reflects the article relevance. The PageRank dependence on $K$ is well
described by a power law $P(K) \propto 1/K^\beta$ (see Fig.~\ref{fig2}) with
$\beta \approx 0.9$ that is consistent with the relation
$\beta=1/(\mu_{in}-1)$ corresponding to the proportionality of PageRank
to its in-degree $w_{in}$ \cite{meyer}.
\begin{figure}
\begin{center}
\includegraphics[clip=true,width=8.2cm]{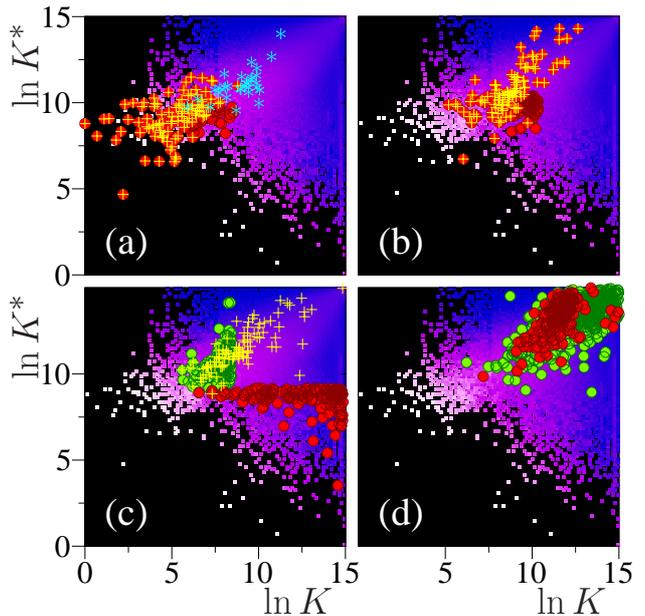}
\vglue -0.1cm
\caption{(Color online) 
Density distribution $W(K,K^*)= d N_i/dK dK^*$ 
of Wikipedia articles in the plane
of PageRank and CheiRank indexes $(\ln K, \ln K^*)$ shown by color
with blue for minimum  and white for maximum (black for zero);
(a) red points show top 100 countries from 2DRank,
yellow pluses show top 100 countries from SJR country documentation 
ranking for 1996-2008 \cite{country}, 
cyan stars mark 30 Dow-Jones companies \cite{dowjones};
(b) red points show top 100 universities from 2DRank, yellow pluses
show top 100 universities from Shanghai ranking of 2009 \cite{shanghai};
(c) green/red points show top 100 personalities from PageRank/CheiRank,
yellow pluses show top 100 personalities from \cite{hart};
(d) green and red points show ranks of 754 physicists, red points mark 193
Nobel laureates.
}
\label{fig3}
\end{center}
\end{figure}

In addition to the PageRank, following the approach 
to Linux Kernel procedure call
network proposed recently by Chepelianskii \cite{alik}, we also
consider the ranking of articles obtained from the 
conjugated Google matrix $G^*$.
This matrix is 
built from the adjacency matrix where all link directions 
are inverted to opposite. After this inversion the Google matrix
$G^*$ is consructed in a usual way
using $\alpha=0.85$. 
The eigenvector $P^*(i)$ of $G^*$
at $\lambda=1$, introduced in \cite{alik}, 
allows to {\it chercher l'information}
in a new way and we call it  CheiRank. 
English spelling of this term sounds as Russian phrase 
which translation is ``whose rank''.
CheiRank gives additional 
ranking of articles in 
decreasing order of $P^*(i)$ with rank index $K^*(i)$.
Our data, shown in Fig.~\ref{fig2}, 
give a power law dependence $P^* \propto 1/{K^*}^\beta$
with $\beta \approx 0.6$ corresponding to the relation
$\beta=1/(\mu_{out}-1)$ similar to the one found for the PageRank.
Both ranks are normalized to unity.

\begin{figure}
\begin{center}
\includegraphics[clip=true,width=4.0cm]{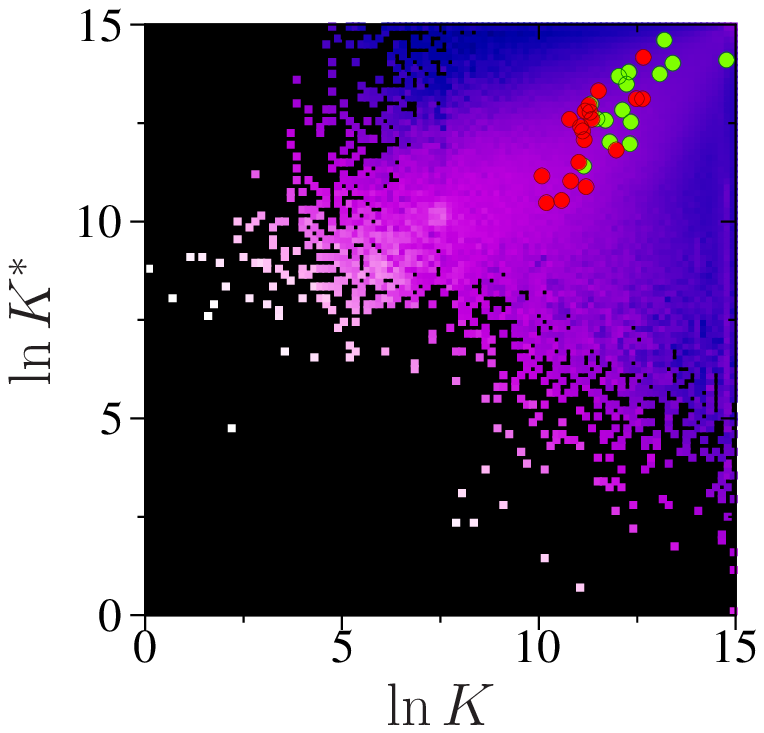}
\includegraphics[clip=true,width=4.0cm]{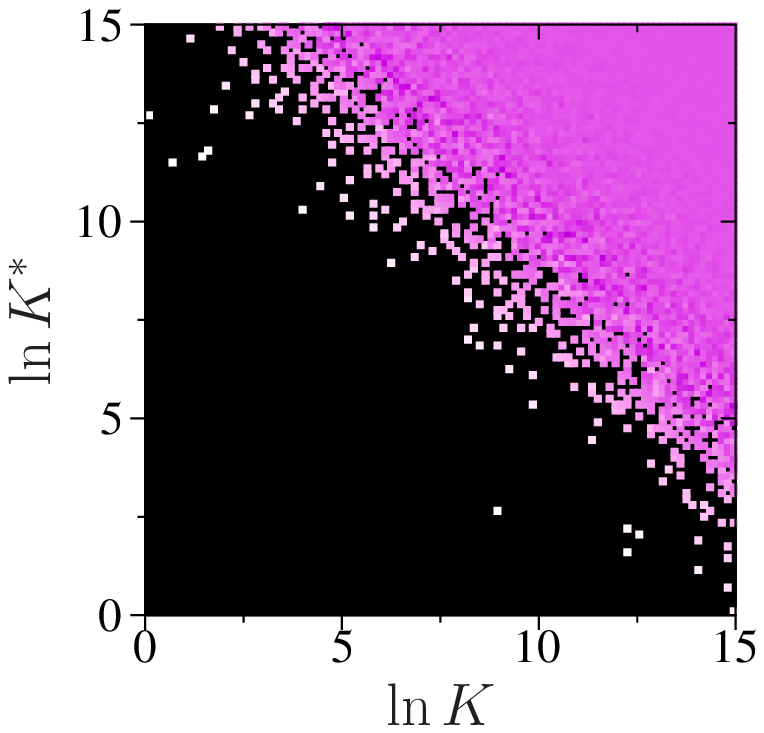}
\vglue -0.1cm
\caption{(Color online) Left panel shows
density distribution of Wikipedia articles
$W(K,K^*) = d N_i/dKdK^*$ in the plane 
of PageRank and CheiRank indexes $(\ln K, \ln K^*)$ as in Fig.~\ref{fig3},
34 green and red points show distribution of chess players,
world champions are marked in red. 
Right panel shows density distribution
of $N=3282257$ articles obtained with independent 
probability distributions generated by $P$ and $P^*$
given by the dependence from Fig.~\ref{fig2}.
Color choice is as in Fig.~\ref{fig3}. 
}
\label{fig4}
\end{center}
\end{figure}

Now all articles (or nodes) can be ordered in decreasing 
monotonic order
of probability of PageRank $P(i)$ or in decreasing monotonic order
of probability of CheiRank $P^*(i)$. In this  way the
ranking of nodes becomes two-dimensional so that
each node $i$ has both CheiRank $K^*(i)$ and PageRank $K(i)$
(see Figs.~\ref{fig3},\ref{fig4}).
Such a ranking on 2D plane $(K,K^*)$
takes into account a combined effect of ingoing 
(for PageRank) and outgoing (for CheiRank) links.
The usual PageRank is obtained by projection of all nodes
on horizontal axis of $K$ and the ranking 
only by outgoing links via CheiRank
is obtained by projecting all nodes on the vertical axis of $K^*$.
The data show that there are no articles which 
have small $K$ and $K^*$ values simultaneously.
We will discuss the rich properties of 2D ranking below in more detail.

\begin{figure}
\begin{center}
\includegraphics[clip=true,width=8.2cm]{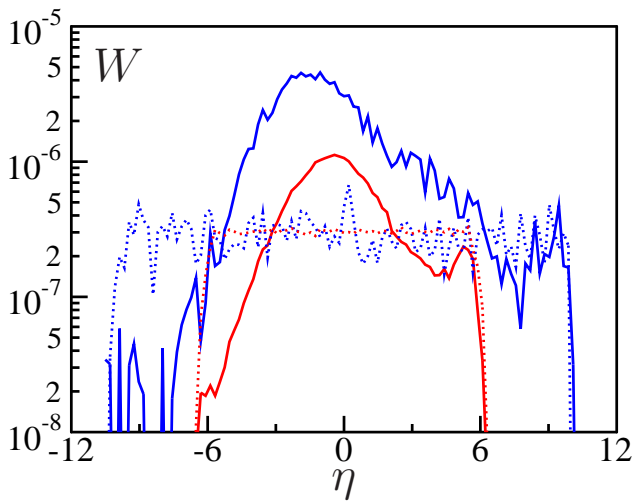}
\vglue -0.1cm
\caption{(Color online) 
Panel  shows $W(K,K^*)$ density 
dependence  from Fig.~\ref{fig3} on a parametric variable $\eta$
which parametrizes
the straight line $\ln K = x_{0}+\eta/2 $, 
$\ln K^* =  x_{0}-\eta/2$ with $x_{0} = 10$ (blue curve)),
$12$ (red curve); dotted curves show densities 
from independent probabilities, of right panel of Fig.~\ref{fig4},
taken at the same $\eta-$lines.
}
\label{fig5}
\end{center}
\end{figure}

\begin{figure}
\begin{center}
\includegraphics[clip=true,width=8.0cm]{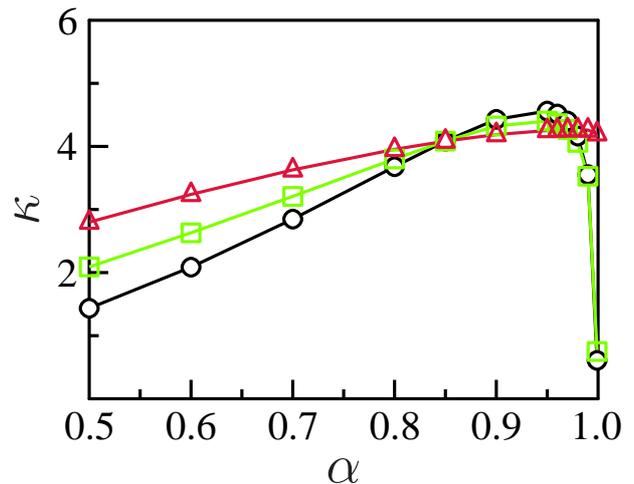}
\vglue -0.1cm
\caption{(Color online)
Dependence of correlator $\kappa$ on $\alpha$.
Black circles, green squares and 
red triangles correspond to cases ($\alpha=\alpha^{*}$), (fixed $\alpha=0.85$
and  $0.5 \leq \alpha^{*} \le 1$) and
(fixed $\alpha^{*}=0.85$ and $0.5 \leq \alpha \le 1$), respectively.
In the main part of the paper we use only the case $\alpha=\alpha^*=0.85$.
}
\label{fig6}
\end{center}
\end{figure}

While PageRank characterizes a degree of knowledge and popularity of
a given site $i$, CheiRank highlights its  communication, 
influence and connectivity degree. 
These ranks have certain analogy to authorities and hubs appearing in
the HITS algorithm \cite{hits} but the HITS is query dependent while the
rank probabilities $P(i)$ and $P^*(i)$ classify all sites of the network.
With these two qualitatively different
characteristics the ranking  of sites becomes two-dimensional (2D).

The density distribution $W(K,K^*)= d N_i/dK dK^*$ of Wikipedia
articles $N_i$ in the plane $(\ln K, \ln K^*)$ 
is shown in Figs.~\ref{fig3},\ref{fig4} ($\sum_i W(K,K^*)  =1$).
Density distribution of Wikipedia articles
$W (K,K^*) = d N_i/dK dK^* $  
is computed over equidistant grid in plane $(\ln K, \ln K^*)$
with $100 \times 100$ cells, color shows average 
value of $W$ in each cell, the normalization condition is
$\sum_{K,K^*} W(K,K^*)=1$.  

The densities 
obtained from the product of independent probabilities
of $P$ and $P^*$ generated by the distributions of Fig.~\ref{fig2}
give very different density $W(K,K^*)$
with $W$ being homogeneous along lines $\ln K^* = - \ln K +const$. 
At large values of $K, K^*$ the distribution $W(K,K^*)$, obtained 
from independent probabilities, is constant, apart from small fluctuations,
since the measure is given by $dK dK^*$ is homogeneous
(see dotted curves in Fig.~\ref{fig5} and right panel in Fig.~\ref{fig4}).
However, at small values of $K, K^*$
the cells, which size decreases with $K, K^*$, 
start to have small number of articles per cell $(\lesssim 1)$ and 
fluctuations give nonhomogeneous behavior of $W(K,K^*)$.

In contrast to the Linux network \cite{alik},
the Wikipedia network has a maximum of density along the
line $\ln K^* \approx 5 + (2 \ln K)/3$ that shows a strong correlation
between $P$ and $P^*$ (see Figs.~\ref{fig3},\ref{fig4}).
Indeed, for Wikipedia we find that the  correlator $\kappa$ between
PageRank and CheiRank probabilities $P(i)$ and $P^*(i)$ is rather high:
\begin{equation}
  \kappa = N \sum_i P(i)P^*(i)-1 = 4.08 \; .
\label{eq2}
\end{equation}
This value is much larger than 
for the Linux network where $\kappa \approx -0.05$. 
Due to these correlations 
the distribution $W(K,K^*)$ is absolutely different from
the one given by independent product probabilities 
for PageRank index $K$ and CheiRank index $K^*$ 
which is homogeneous along lines $\ln (K^*)+\ln( K) =const$ 
(see Fig.~\ref{fig4} (right panel) and Fig~\ref{fig5}). 
Indeed, the real density of articles
$W(K,K^*)$,  taken 
along a given $\eta-$line in the plane $(\ln K, \ln K^*)$, is log-normal and 
has a Gaussian form with a certain width as it is shown in Fig.~\ref{fig5}.

The value of correlator $\kappa$ for Wikipedia is comparable with the one 
of Cambridge University network ($\kappa=3.79$ \cite{alik})
but the probability distributions are different. 
Dependencies of the correlator $\kappa$ on  damping
parameters $\alpha$ and $\alpha^*$  are shown in Fig.~\ref{fig6}
(here $\alpha^*$ is the damping parameter for the Google matrix $G^*$
with inverted direction of links). For all realistic values of
$\alpha$ and $\alpha^*$ the correlator remains larger than unity
showing that the correlation between CheiRank and PageRank is 
a robust feature of the Wikipedia hyperlink network.

From a physical view point a positive value of correlator
$\kappa$ marks the tendency to have 
more outgoing and ingoing links between certain sets of nodes.
This property is clearly visible for Cambridge
University network,  as discussed in \cite{alik},
and for Wikipedia network discussed here.
On the other side small or negative values of $\kappa$
correspond to anti-correlation between ingoing and outgoing links,
that looks like an effective repulsion between them.
Such a situation appears in the Linux Kernel network \cite{alik}.
Future more detailed studies of  this correlator
should be done for other types of networks
to understand its properties in a better way.

\section{Ranking of selected categories}

The difference between PageRank and CheiRank is clearly seen
from the names of articles with highest rank
(detailed data for all ranks and all categories considered are
given in Appendix and \cite{2drank}). At the top of PageRank we have 
1. {\it United States}, 2. {\it United Kingdom}, 3. {\it France} 
while for CheiRank we find
1. {\it Portal:Contents/Outline of knowledge/Geography and places},
2. {\it List of state leaders by year}, 
3. {\it Portal:Contents/Index/Geography and places}. Clearly PageRank
selects first articles on a broadly known subject 
with a large number of ingoing links
while CheiRank selects first highly communicative articles with
many outgoing links. Since the articles are distributed in 2D
they can be ranked in various ways corresponding to projection of 2D set
on a line. The horizontal and vertical lines correspond 
to PageRank and CheiRank. 

\begin{figure}
\begin{center}
\includegraphics[clip=true,width=8.2cm]{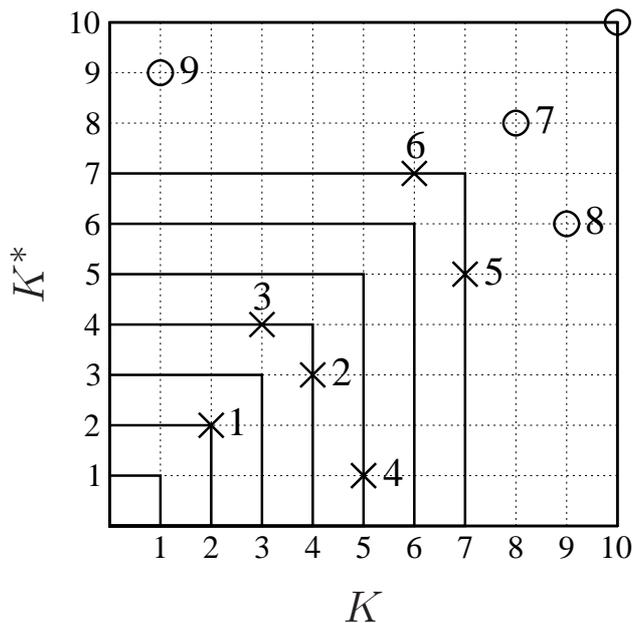}
\vglue -0.1cm
\caption{(Color online) 
A toy example illustrating the functioning of 2DRank 
algorithm of node ranking  in ${K,K^*}$ plane: 
square size $k \times k$ is
regularly increase in size $k \rightarrow k+1$,
nodes appearing on edges of this square at each step
are included in the listing $K_2$ of 2DRank (crosses),
first on right edge, then on top edge;
nodes outside of the square (circles) are included in the listing
$K_2$ at later stage. Numbers near symbols give
$K_2$ values of 2DRank. 
}
\label{fig7}
\end{center}
\end{figure}

We also introduce 2DRank $K_2(i)$ to characterize mixed effect of 
CheiRank and PageRank.  The list of  $K_2(i)$ is constructed by  
increasing $K \rightarrow K+1$ 
and increasing 2DRank index 
$K_2(i)$ by one if a new entry is present in the list of
first $K^*<K$ entries of CheiRank, then the one unit step is done in
$K^*$ and $K_2$ is increased by one if the new entry is
present in the list of first $K<K^*$ entries of CheiRank.
More formally, 2DRank $K_2(i)$ gives the ordering 
of the sequence of sites, that $\;$ appear
inside  $\;$ the squares  $\;$
\newline
$\left[ 1, 1; \;K = k, K^{\ast} = k; \; \-... \right]$ when one runs
progressively from $k = 1$ to $N$. In fact, at each step
$k \rightarrow k + 1$ there are tree possibilities: 
(i) no new sites on two edges of square, 
(ii) only one site is on these two edges 
and it is added in the listing of $K_2(i)$
and (iii) two  sites are on the edges and both are added 
in the listing $K_2(i)$, first with $K > K^{\ast}$ 
and second with $K < K^{\ast}$. For (iii) the choice of order
of addition in the list $K_2(i)$ affects only some pairs of  
neighboring sites and does
not change the main structure of ordering.
This 2DRank algorithm is illustrated in Fig.~\ref{fig7}
by a toy example.

Wikipedia articles with highest 2DRank are 1. {\it India},
2. {\it Singapore}, 3. {\it Pakistan}. Thus, these articles
are most known/popular and most communicative
at the same time. We note that the ranking of Wikipedia articles 
by PageRank and HITS authorities 
has been discussed in a literature (see e.g.  \cite{bellomi,lopes})
but 2D analysis was never done before.

To understand the properties of three ranks $K, K_2, K^*$
in a better way we consider the three main categories of 
articles about countries, universities and personalities.
The locations of 100 top internationally established ranks, 
taken according to
\cite{country,shanghai,hart} respectively, are shown in 
the Wikipedia rank plane $(\ln K, \ln K^*)$ in Fig.~\ref{fig3}a,b,c.
In average the points are distributed along the band of 
maximal density of $W$ in this plane. The same is valid for 
Dow-Jones companies (Fig.~\ref{fig3}a) 
and  physicists (Fig.~\ref{fig3}d)
taken from the Wikipedia List of physicists (with few additions).
We also show 100 top countries (Fig.~\ref{fig3}a),
100 top universities (Fig.~\ref{fig3}b) according to 2DRank
and 100 top personalities according 
to PageRank and CheiRank (Fig.~\ref{fig3}c).
Distribution of chess players is shown in left panel of Fig.~\ref{fig4}.

According to PageRank and 2DRank the chosen categories are ordered 
as following:
countries (category top rank values 
$K=1$, $K_2=1$, $K^*=107$), 
universities ($K=180$, $K_2=5$, $K^*=836$), 
personalities ($K=268$, $K_2=233$, $K^*=33$), 
Dow-Jones companies ($K=366$, $K_2=1082$, $K^*=14298$),
physicists ($K=505$, $K_2=1202$, $K^*=7622$) and 
chess players ($K=23903$, $K_2=5354$, $K^*=35670$). 
Such an ordering of these categories
corresponds to a natural influence proportional to an 
effective size of categories (e.g. countries are larger than universities,
which are larger than personalities etc.). 
Pleasantly enough, from the point of Wikipedia ranking,
universities are much more influential
than Dow-Jones companies. At the same time it is interesting 
that for chosen categories the communicative power measured by CheiRank
is most high for personalities.

\begin{figure}
\begin{center}
\includegraphics[clip=true,width=8.2cm]{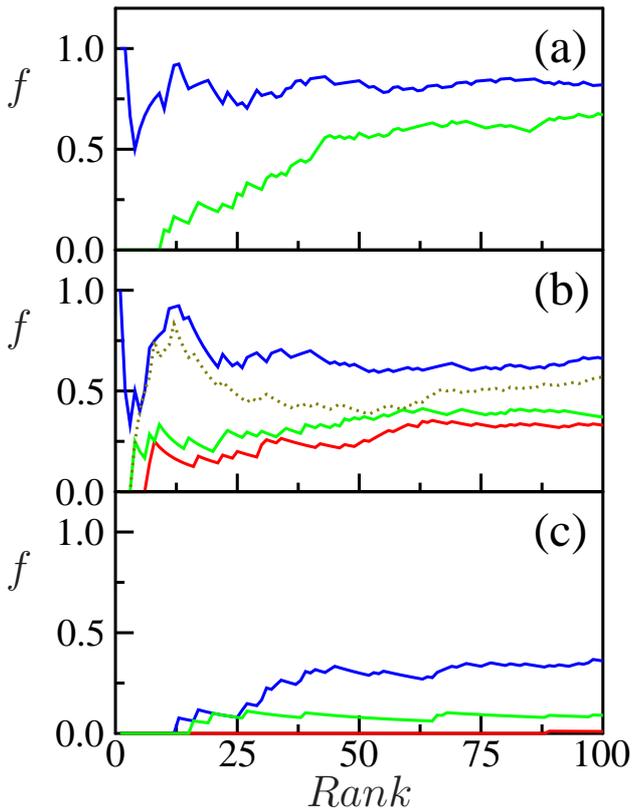}
\vglue -0.1cm
\caption{(Color online) 
Dependence of overlap fraction $f$ of top Wikipedia articles 
for various categories on selected 
rank $K_s$ taken from different internationally established sources:
(a) SJR country documentation rank for 1996-2008 \cite{country};
(b) Shanghai university rank of 2009 \cite{shanghai};
(c) Hart's personality rank \cite{hart}.
Colors mark
Wikipedia PageRank $K$ (blue), 2DRank $K_2$ (green) 
and CheiRank $K^*$ (red). In (a) the order of top 100 countries is 
the same for 2DRank and CheiRank; in (b) the gray dotted curve
shows also $f$ for the rank given by advanced Google search of a word
{\it university} in the Wikipedia English domain at May 19, 2010.
}
\label{fig8}
\end{center}
\end{figure}

\begin{figure}
\begin{center}
\includegraphics[clip=true,width=8.2cm]{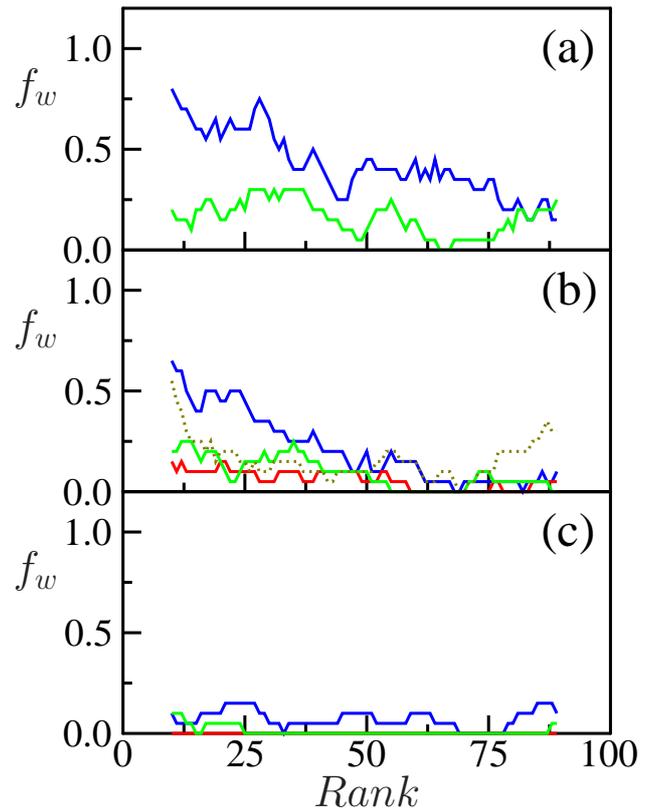}
\vglue -0.1cm
\caption{(Color online) 
Dependence of window overlap fraction $f_w$ 
for top Wikipedia articles on internationally established ranks
from Fig.~\ref{fig8} respectively for panels (a), (b), (c)
(points are placed at the middle of window interval,
window size is 20 nodes).
}
\label{fig9}
\end{center}
\end{figure}

We discuss the main features of three ranks $K, K_2, K^*$
starting from countries. The first three countries
are 1. {\it United States}, 2. {\it United Kingdom}, 3. {\it France}
for PageRank and
1. {\it India}, 2. {\it Singapore}, 3. {\it Pakistan}
for 2DRank and CheiRank. 
To determine in a quantitative way how 
accurate is this ranking we introduce the overlap fraction $f$
defined as the relative number of same entries 
inside the first $K_s$ entries of rank
of selected countries, as stated in  \cite{country}, 
and first $K_s$ countries according to
$K, K_2, K^*$ ranking. The dependence $f(K_s)$ is shown in Fig.~\ref{fig8}a.
We see that the Wikipedia PageRank reproduces in average about 80\%
of top countries selected by \cite{country}. CheiRank and 2DRank 
do not reproduce the initial values of rank \cite{country}
but reach around 70\% at maximum $K_s=100$. Indeed, 
CheiRank and 2DRank place at the top countries
which are not so authoritative as United States but which
are strongly connective due to historical 
(e.g. Egypt is at position 13) or tourist reasons
(e.g. Thailand at , Malaysia at 7).
Definitely the first triplet of PageRank gives western countries 
which are better known and historically more powerful.
However, the first triplet of CheiRank and 2DRank
shows that the communicative power of other countries, even 
such small as {\it Singapure}, dominates  on the pages of Wikipedia.
This may be a sign of future changes. Let us also note that
{\it People's Republic of China} is only at 26 position in 2DRank
that can be attributed to non-English origin of the 
country and political separation of historical {\it China}
(its SJR rank is 4).

According to Wikipedia the top universities are
1.{\it Harvard University}, 2. {\it University of Oxford},
3. {\it University of Cambridge} in PageRank;
1. {\it Columbia University}, 2. {\it University of Florida},
3. {\it Florida State University} in 2DRank and CheiRank.
The fraction $f(K_s)$ for overlap with
Shanghai university ranking \cite{shanghai}
is shown in Fig~\ref{fig8}b. The Wikipedia PageRank  
reproduces in average around 70\% of ranking \cite{shanghai}
that is about 10\% higher than gives the advanced Google search.
CheiRank and 2DRank at maximum give around 25\% and 35\%.
These ranks highlight connectivity degree of universities
that leads to appearance of significant
number of  arts, religious and military specialized colleges 
(12\% and 13\% respectively for CheiRank and 2DRank)
while PageRank has only 1\% of them. 
CheiRank and 2DRank introduce also
a larger number of relatively small universities.
We argue that such colleges and universities keeps links to their 
alumni in a significantly better way that increases their ranks.

For personalities the Wikipedia ranking gives
1. {\it Napoleon I of France}, 2. {\it George W. Bush},
3. {\it Elizabeth II of the United Kingdom} for PageRank;
1.{\it Michael Jackson}, 2. {\it Frank Lloyd Wright},
3. {\it David Bowie} for 2DRank;
1. {\it Kasey S. Pipes}, 2. {\it Roger Calmel},
3. {\it Yury G. Chernavsky} for CheiRank.
The overlap fraction $f(K_s)$ for Hart's personality ranking \cite{hart}
is shown in Fig.~\ref{fig8}c. Even for the PageRank
it is at maximum 35\%  being around 10\% for 2DRank and
almost zero for CheiRank. We attribute this to a 
very broad distribution of personalities in 2D plane (Fig.~\ref{fig3}c)
and a large variety of human activities which 
we classify by 5 main categories:
politics, religion, arts, science, sport. 
For top 100 PageRank personalities we have
for these categories: 58, 10, 17, 15, 0 respectively.
Clearly PageRank overestimates the significance of politicians.
For 2DRank we find respectively 24, 5,  62, 7, 2. Thus this rank 
highlights artistic sides of human activity.
For CheiRank we have 15, 1, 52, 16, 16 so that the dominant
contribution comes from arts, science and sport. The interesting property 
of this rank is that it selects many composers, singers, writers, actors.
As an interesting feature of CheiRank we note
that among scientists it selects those who are not so much known to
a broad public but who  discovered new objects, e.g. 
George Lyell who discovered many Australian butterflies  
or  Nikolai Chernykh who discovered many asteroids. 
CheiRank also selects persons active in several categories of human activity.

Of course, the overlap fraction $f$ gives only
the integral overlap between Wikipedia ranking and 
the selected internationally recognized ranking.
It is possible to get more refined comparison taking the
window fraction $f_w$ of relative number of
names being the same inside a certain window
of fixed size. We present such type of data 
for $f_w$ in Fig.~\ref{fig9}
for a window size of 20 nodes.
For countries and universities the values of $f_w$ are rather
high for top 40 ranks but for higher ranks there is a gradual 
decrease of $f_w$. We attribute this to increasing fluctuations
at large values of rank where each item can move 
in a significant interval. For personalities the window fraction $f_w$
remains rather low. We attribute this to a very broad variety
of human activity as discussed above.

When a human activity is fixed in a more precise way then 
the Wikipedia ranking gives a rather reliable ordering.
For example, for 754 physicists (see Fig.~\ref{fig3}d) 
we find at the top: 1. {\it Aristotle}, 2. {\it Albert Einstein},
3. {\it Isaac Newton} from PageRank;
1. {\it Albert Einstein}, 2. {\it Nikola Tesla},
3. {\it Benjamin Franklin} from 2DRank;
1. {\it Hubert Reeves}, 2. {\it Shen Kuo},
3. {\it Stephen Hawking} from CheiRank.
It is clear that PageRank gives most known,
2DRank gives most known and active in other areas,
CheiRank gives those who are known and contribute to popularization
of science. Indeed, e.g. {\it Hubert Reeves} and
{\it Stephen Hawking} are very well known for their popularization
of physics that increases their communicative power and 
place them at the top of CheiRank. {\it Shen Kuo} obtained recognized
results in an enormous variety of fields of science that 
leads to the second top position in CheiRank even if his activity
was about thousand years ago.

\begin{figure}
\begin{center}
\includegraphics[clip=true,width=8.2cm]{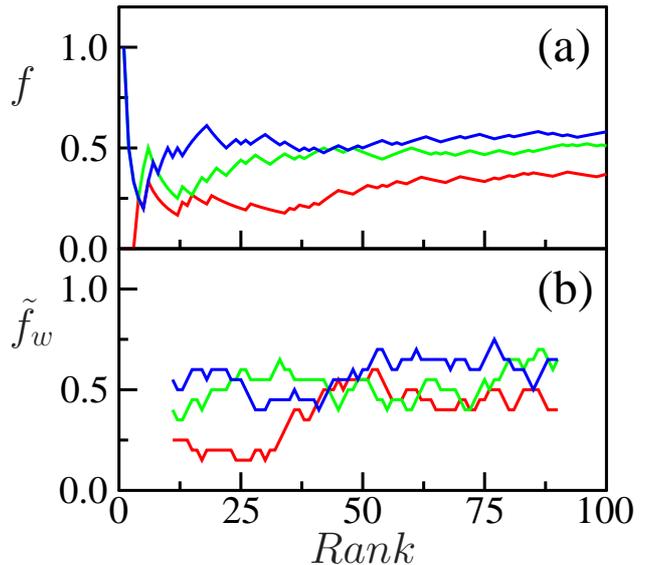}
\vglue -0.1cm
\caption{(Color online) 
(a) Overlap fraction $f$ of Nobel laureate physicists  
as a function of rank
of physicists via Wikipedia articles;
(b) window overlap fraction $f_w$ 
for the data of panel (a), 
points are placed at the middle of window interval, window size is 20 nodes.
Colors mark Wikipedia PageRank $K$ (blue), 2DRank $K_2$ (green) 
and CheiRank $K^*$ (red) which ranks are plotted on horizontal axis.
}
\label{fig10}
\end{center}
\end{figure}

For physicists, who lived at the time of Nobel prize
and could get it, we can determine the overlap fraction $f$
as a relative number of Nobel laureates at a given rank
value $K, K_2, K^*$ (see curves in Fig.~\ref{fig10}a).
In certain aspects, this definition of $f$ for physicists is 
different from the case of categories in Fig.~\ref{fig8}
since here Nobel laureates are in the same listing of Wikipedia ranks
of physicists
while in Fig.~\ref{fig8} the rank $K_s$ is taken from independent 
classification source.  
The data for Nobel laureates in Fig.~\ref{fig10}a give high 
average values of $f \approx 0.5$ for PageRank and 2DRank,
and $f \approx 0.25$ for CheiRank.
We note that the number of Nobel prizes is not so large
and even very notable physicists remained without it
(e.g. Thomas Edison, Nikola Tesla, Alexander Graham Bell
are those from the top of PageRank who remained without the prize). 
Hence, the prediction level of $f\approx 0.5$
can be considered as rather high.

The window overlap fraction $f_w$ for Nobel laureates is shown in
Fig.~\ref{fig10}b, here inside the window
$f_w$ gives the relative fraction of Nobel laureates 
without coincidence of concrete names 
(Nobel laureates form a sublisting of listing of physicists;
note that in Fig.~\ref{fig9} we look  overlap of same 
article titles from a pair of different listings).
Even at large rank values $f_w$ remains on the level 
of 40-50\% that is twice higher than the average fraction 
of Nobel laureates in the list of physicists 
$<f> = 193/754 \approx 0.25$.
This shows that Wikipedia ranking have high correlations with the 
list selected by the Nobel committee.

The ranking  of 193 Nobel laureate physicists give
1.  {\it Albert Einstein}, 2. {\it Enrico Fermi},
3. {\it Richard Feynman} from PageRank;
1. {\it Albert Einstein}, 2. {\it Richard Feynman},
3. {\it Wer\-ner Heisenberg} from 2DRank and CheiRank.
The communication power is well visible if to consider
certain specific names, e.g. {\it Abdus Salam}
has PageRank 20 and CheiRank 5. 
It is clear that the communicative power
of this article is strongly enhanced due to
existence of 
the Abdus Salam International Centre for Theoretical Physics.

Thus Wikipedia ranking can be applied to various categories giving
rather good results without any fitting. 
For example, for Dow-Jones companies we have
1. {\it Microsoft}, 2. {\it IBM}, 3. {\it The Walt Disney Company}
from PageRank and
1. {\it Cisco Systems}, 2. {\it The Walt Disney Company},
3. {\it Microsoft} from CheiRank. 
CheiRank correctly selects 
{\it Cisco Systems},  the worldwide leader in networking,
at the top of communication power of Dow-Jones listing. 

For chess players we find:
1. {\it Garry Kasparov}, 2. {\it Bobby Fischer},
3. {\it Alexander Alekhine} from PageRank $K$;
1. {\it Bobby Fischer}, 2. {\it Alexander Alekhine}, 
3. {\it Emanuel Lasker} from 2DRank $K_2$ and
1. {\it Bobby Fischer}, 2. {\it Alexander Alekhine},
3. {\it Wilhelm Steinitz} from CheiRank $K^*$. 
Thus world chess champions,
including most strong, 
appear at the top positions. 
Indeed, according to FIDE \cite{chess} {\it Garry Kasparov}
is ranked number 1 most times.
Clearly a lot 
of hot stories around {\it Bobby Fischer} increased his 2DRank and CheiRank.

\section{Discussion}

On the basis of presented results we conclude that the ranking of Wikipedia
articles allows to rank human knowledge in a rather reliable way.
The 2D ranking highlights the properties of articles in a new
rich and  fruitful manner. We think that such type of 2D ranking 
will find further useful applications for various complex networks 
including the WWW. 

It would be also interesting to
apply the 2D ranking to citation network of Physical Review where PageRank 
is known to give a reliable ranking of physicists 
\cite{redner,fortunato}. We suppose that 
for this network CheiRank will allow
to  select notable reviews which played an important role 
in the propagation of knowledge between various fields of physics.
It is possible that combination of 2D ranking 
with other methods of various entities ranking of Wikipedia
can be also useful (see e.g. \cite{zaragoza}).

Wikipedia becomes the largest library of human knowledge.
``The Library exists {\it ab aeterno}'' 
declared Jorge Luis Borges \cite{borges}. Thus, the ranking of 
this enormous amount of knowledge
becomes a formidable challenge and we think that the 
2D ranking will play for this task a useful and important role.  

\renewcommand{\theequation}{A-\arabic{equation}}
  \setcounter{equation}{0}  
\renewcommand{\thefigure}{A-\arabic{figure}}
  \setcounter{figure}{0}  
\section{Appendix}

Below we give the listings of Wikipedia articles 
in their rank order according to the three ranks discussed in the paper
for the categories discussed. More detailed data are available at 
\cite{2drank}.

L1) Top 20  Wikipedia articles:

PageRank: 1. United States, 2. United Kingdom, 3. France,
4. Germany, 5. England, 6. Canada, 7. World War II, 8. Australia,
9. India, 10. Japan, 11. English Language, 12.Italy,
13.  Association football, 14. 2007, 15. London, 16. Poland,
17. Animal, 18. List of sovereign states, 19. 2008, 20. New York City.

2DRank:   1. India, 2. Singapore, 3. Pakistan, 4. Brazil,
5. Columbia University, 6. Thailand, 7. Amphibian, 8. Chile, 9. Fossil,
10. Malaysia, 11. Yorkshire, 12. New York Yankees, 13. Virginia,
14. Iceland, 15. California, 16. England, 17. New Jersey,
18. Michigan, 19. Hong Kong, 20. Municipalities of Switzerland.

CheiRank: 1. Portal:Contents/Outline of knowledge/\-Geography and places,
2. List of state leaders by year, 3. Portal:Contents/Index/\-Geography and places, 
4. Lists of country-related topics, 5. Portal:Mathematics/Index, 
6. List of United Kingdom locations, 7. Navy Office of Community Outreach,
8. Portal:Spaceflight/Status, 9. List of Tachinidae genera and species,
10. Village Development Committee, 11. Tachinidae, 12. Navy Weeks,
13. Lists of birds by region, 14.  Outline of Africa, 15. List of Bulbophyllum species, 
16. Bulbophyllum, 17. Index of Thailand-related articles,
18. Portal:Middle-earth/Pages, 19. Portal:Trains/Anniversaries,
20. Portal:Contents/Outline of knowledge/History and events.

L2) Top 20 articles in category countries:

PageRank: 1. United States, 2. United Kingdom, 3. France,
4. Germany, 5. Canada, 6. Australia, 7. India, 8. Japan,
9. Italy, 10. Poland, 11. Spain, 12. Russia, 13. Netherlands,
14. Brazil, 15. New Zealand, 16. Sweden, 17. Romania, 18. Switzerland,
19. Mexico, 20. Norway.

2DRank and CheiRank give the same order: 1. India, 2. Singapore, 3. Pakistan, 4. Brazil, 5. Thailand,
6. Chile, 7. Malaysia, 8. Iceland, 9. Hong Kong. 10. Canada,
11. Peru, 12. Afghanistan, 13. Egypt, 14. Armenia, 15. Argentina,
16. United Kingdom, 17. Colombia, 18. Finland, 19. Benin, 20. Bolivia.

United States are at rank $41$. Top 100 countries have the same order
in 2DRank and CheiRank. We attribute this to a rather sparse
distribution in CheiRank index between countries.
SJR country documentation rank for 1996-2008 
is available at \cite{country}.

L3) Top 20 articles in category universities:

PageRank: 1. Harvard University, 2. University of Oxford, 
3. University of Cambridge, 4. Columbia University, 
5. Yale University, 6. Massachusetts Institute of Technology, 
7. Stanford University, 8. University of California, Berkeley,
9. Princeton University, 10. Cornell University, 11. University of Chicago,
12. University of Michigan, 13. University of California, Los Angeles,
14. University of Pennsylvania, 15. New York University,
16. University of Texas at Austin, 17. University of Toronto,
18. University of Southern California, 19. University of Virginia,
20. University of Florida.

2DRank: 1. Columbia University, 2. University of\- Florida,
3. Florida State University, 4. University of California, Berkeley,
5. Northwestern University, 6. Brown University, 
7. University of Southern California, 8. Carnegie Mellon University, 
9. Massachusetts Institute of Technology, 
10. University of Michigan, 11. Georgetown University,
12. Juilliard School, 13. University of Pittsburgh, 
14. Amherst College, 15. Cornell University,
16.  Durham University, 17. Rutgers University,
18. Monash University, 19. The University of Western Ontario,
20. University of Toronto.

CheiRank: 1. Columbia University, 2. University of Florida,
3. Florida State University, 4. Brooklyn College,
5. Amherst College, 6. The University of Western Ontario,
7. University of Sheffield, 8. University of California, Berkeley,
9. Northwestern University, 10. Northeastern University, 11. Brown University, 
12. Queen's University of Belfast (Northern Ireland Parliament constituency), 
13. Fairfield University, 14. University of Southern California, 
15. Carnegie Mellon University, 16. Grambling State University,
17. Massachusetts Institute of Technology, 18. University of Michigan,
19. Georgetown University, 20. University of Pittsburgh.

Rank given by advanced Google search in Wikipedia English domain
at May 19, 2010:   1. Princeton University, 2. Columbia University,
3. University of Oxford, 4. Stanford University,
5. Harvard University, 6. University of Cambridge,
7. Cornell University, 8. University of California, Berkeley, 
9. Yale University, 10. University of Virginia,
11. Northwestern University, 12. University of\- Chicago,
13. Brown University, 14. University of Michigan,
15. Rutgers University, 16. University of Washington, 
17. Indiana University, 18. University of Minnesota,
19. Howard University, 20. Leiden University.

Shanghai university rank of 2009 is available at \cite{shanghai}.

L4) Top 20 articles in category personalities:

PageRank: 1. Napoleon I of France, 2. George W. Bush, 
3. Elizabeth II of the United Kingdom, 4. William Shakespeare, 
5. Carl Linnaeus, 6. Adolf Hitler,
7.  Aristotle, 8. Bill Clinton, 9. Franklin D. Roosevelt, 10. Ronald Reagan,
11. Barack Obama, 12. Richard Nixon, 13. George Washington, 14. Joseph Stalin,
15. Abraham Lincoln, 16. John F. Kennedy, 17. Muhammad, 18. Winston Churchill,
19. Henry VIII of England, 20. Alexander the Great.

2DRank: 1. Michael Jackson, 2. Frank Lloyd Wright, 3. David Bowie,
4. Hillary Rodham Clinton, 5. Charles Darwin, 6. Stephen King,
7. Richard Nixon, 8. Isaac Asimov, 9. Frank Sinatra, 10. Elvis Presley,
11. Edward Elgar, 12. Stephen Sondheim, 13. Agatha Christie,
14. Pope John Paul II, 15. Robert A. Heinlein, 16. Adolf Hitler,
17. Madonna (entertainer), 18. Ozzy Osbourne, 19. John McCain,
20. Jesus.

CheiRank: 1. Kasey S. Pipes, 2. Roger Calmel, 3. Yury G. Chernavsky,
4. Josh Billings (pitcher), 5. George Lyell, 6. Landon Donovan,
7. Marilyn C. Solvay, 8. Matt Kelley, 9. Johann Georg Hagen,
10. Chikage Oogi, 11. Bobbie Vaile, 12. Rosie Malek-Yonan,
13. Blythe McGarvie, 14. Djoko Hardono, 15. Cristina Bella,
16. Sid Deuce, 17. Joey Hamilton, 18. Kiki Dee,
19. Carlos Francis, 20. Percy Jewett Burrell.

Hart's personality rank is available at \cite{hart} and at\\ 
http://www.adherents.com/adh\_influ.html .

L5) Top  20 articles in category physicists:

PageRank: 1. Aristotle, 2. Albert Einstein, 3. Isaac Newton,
4. Thomas Edison, 5. Benjamin Franklin, 6. Gottfried Leibniz,
7.  Avicenna, 8. Carl Friedrich Gauss, 9. Galileo Galilei,
10.  Nikola Tesla, 11. Andre-Marie Ampere,
12. Michael Faraday, 13. Leonhard Euler, 14. Alexander Graham Bell,
15. James Clerk Maxwell, 16. Archimedes, 17. Blaise Pascal,
18. Stephen Hawking, 19. Enrico Fermi, 20. Johannes Kepler.

2DRank: 1. Albert Einstein, 2. Nikola Tesla,
3. Benjamin Franklin, 4. Avicenna, 5. Isaac Newton,
6. Thomas Edison, 7. Stephen Hawking, 8. Gottfried Leibniz,
9. Richard Feynman, 10. Aristotle, 11. Alhazen (Ibn al-Haytham) — Iraq,
12. Werner Heisenberg, 13. Heinrich Hertz, 14. Johannes Kepler,
15. Galileo Galilei, 16. Shen Kuo, 17. Abu Rayhan Biruni-- Persian,
18. Alexander Graham Bell, 19. Robert Hooke, 20. Michael Faraday.

CheiRank: 1. Hubert Reeves, 2.  Shen Kuo, 3. Stephen Hawking,
4. Nikola Tesla, 5. Albert Einstein, 6. Arthur Stanley Eddington,
7. Richard Feynman, 8. John Joseph Montgomery, 9. Josiah Willard Gibbs,
10. Heinrich Hertz, 11. Benjamin Franklin, 12. Edwin Hall,
13. Avicenna, 14. Isaac Newton, 15. Thomas Edison,
16.  Michio Kaku, 17. Abu Rayhan Biruni-- Persian,
18. Abdul Qadeer Khan, 19. Werner Heisenberg,
20. Gottfried Leibniz.

L6) Top 20 articles in category Nobel laureate physicists
(physicists who got any Nobel prize):

PageRank: 1. Albert Einstein, 2. Enrico Fermi, 3. Richard Feynman,
4. Max Planck, 5. Guglielmo Marconi, 6. Werner Heisenberg, 7. Marie Curie,
8. Niels Bohr, 9. Paul Dirac, 10. J.J.Thomson, 11. Max Born,
12. John Strutt, 3rd Baron Rayleigh, 13. Andrei Sakharov,
14. Pierre Curie, 15. Subrahmanyan Chandrasekhar, 16. Wolfgang Pauli,
17. Lev Landau, 18. Eugene Wigner, 19. Albert Abraham Michelson,
20. Abdus Salam.

2DRank: 1. Albert Einstein, 2. Richard Feynman, 3. Werner Heisenberg,
4. Enrico Fermi, 5. Max Born, 6. Marie Curie, 7. Wolfgang Pauli,
8. Max Planck, 9. Eugene Wigner, 10. Paul Dirac,
11. Guglielmo Marconi, 12. Abdus Salam, 13. Hans Bethe,
14. Andrei Sakharov, 15. Steven Chu, 16. Niels Bohr,
17. J.J.Thomson, 18. Steven Weinberg, 19. Peter Debye,
20. Subrahmanyan Chandrasekhar.

CheiRank: 1. Albert Einstein, 2. Richard Feynman, 3. Werner Heisenberg,
4. Brian David Josephson, 5. Abdus Salam, 6. C.V.Raman, 7. Peter Debye,
8. Enrico Fermi, 9. Wolfgang Pauli, 10. Steven Weinberg,
11. Max Born, 12. Eugene Wigner, 13. Marie Curie,
14. Luis Walter Alvarez, 15. Percy Williams Bridgman,
16. Roy J. Glauber, 17. Max Planck, 18. Paul Dirac,
19. Guglielmo Marconi, 20. Hans Bethe.

L7) Top 30 articles in category chess players:

PageRank: 1. Garry Kasparov, 2. Bobby Fischer, 3. Alexander Alekhine,
4. Anatoly Karpov, 5. Emanuel Lasker, 6. Mikhail Botvinnik,
7. Vladimir Kramnik, 8. Viswanathan Anand,
9. Paul Keres, 10. Boris Spassky, 11. Veselin Topalov,
12. Wilhelm Steinitz, 13. Tigran Petrosian, 14. Max Euwe,
15. David Bronstein, 16. Mikhail Tal, 17. Viktor Korchnoi,
18. Vasily Smyslov, 19. Samuel Reshevsky, 20. Bent Larsen,
21. Jose Raul Capablanca, 22. Boris Gelfand,
23. Gata Kamsky, 24. Alexei Shirov, 25. Mark Taimanov,
26. Magnus Carlsen, 27. Efim Geller,
28. Ruslan Ponomariov, 29. Rustam Kasimdzhanov,
30. Alexander Khalifman.

2DRank: 1. Bobby Fischer, 2. Alexander Alekhine, 3. Emanuel Lasker,
4. Garry Kasparov, 5. Wilhelm Steinitz, 6. Paul Keres,
7. Mikhail Botvinnik, 8. Jose Raul Capablanca, 9. Bent Larsen,
10. Boris Spassky, 11.  Viswanathan Anand, 12. Magnus Carlsen,
13. Vladimir Kramnik, 14. Efim Geller, 15. Samuel Reshevsky, 
16. Anatoly Karpov, 17. Mikhail Tal, 18. Viktor Korchnoi,
19. Max Euwe, 20. Veselin Topalov, 21. Gata Kamsky,
22. Tigran Petrosian, 23. David Bronstein, 24. Rustam Kasimdzhanov,
25. Ruslan Ponomariov, 26. Vasily Smyslov, 27. Alexei Shirov,
28. Boris Gelfand, 29. Wolfgang Unzicker, 30. Mark Taimanov.

CheiRank: 1. Bobby Fischer, 2. Alexander Alekhine, 3. Wilhelm Steinitz,
4. Emanuel Lasker, 5. Garry Kasparov, 6. Paul Keres, 7.  Mikhail Botvinnik,
8. Jose Raul Capablanca, 9. Magnus Carlsen,
10. Bent Larsen, 11. Boris Spassky, 12.  Viswanathan Anand,
13. Vladimir Kramnik, 14. Efim Geller, 15. Samuel Reshevsky,
16. Anatoly Karpov, 17. Mikhail Tal, 18. Viktor Korchnoi,
19. Max Euwe, 20. Veselin Topalov, 21. Gata Kamsky,
22. Tigran Petrosian, 23. David Bronstein, 24. Rustam Kasimdzhanov,
25. Ruslan Ponomariov, 26. Vasily Smyslov, 27. Alexei Shirov,
28. Boris Gelfand, 29. Wolfgang Unzicker, 30.  Mark Taimanov.

L8) Ranking of 30 Dow-Jones companies:

PageRank: 1. Microsoft, 2. IBM, 3. The Walt Disney Company,
4. Intel Corporation, 5. Hewlett-Packard, 6. General Electric,
7. Mcdonald's, 8. Boeing, 9. At\&t, 10. Cisco Systems,
11. DuPont, 12. ExxonMobil, 13. Procter \& Gamble, 14. Bank of America,
15. Verizon Communications, 16. JPMorgan Chase, 17. American Express,
18. Pfizer, 19. The Coca-Cola Company, 20. American Express,
21. Chevron Corporation, 22. 3M, 23. Merck \& Co., 24. The Home Depot,
25. Alcoa, 26. Johnson \& Johnson, 27. Kraft Foods, 28. Caterpillar Inc.,
29. United Technologies Corporation, 30. The Travelers Companies.

2DRank and CheiRank have the same order:
1. Cisco Systems, 2. The Walt Disney Company, 3. Microsoft,
4. Kraft Foods, 5. IBM, 6. At\&t, 7. Hewlett-Packard, 8. Pfizer,
9. Intel Corporation, 10. ExxonMobil, 11. Caterpillar Inc.,
12. DuPont, 13. General Electric, 14. American Express,
15. Johnson \& Johnson, 16.  Boeing, 17. Wal-Mart, 18. Bank of America,
19. Verizon Communications, 20. JPMorgan Chase, 21. Merck \& Co.,
22. The Coca-Cola Company, 23. 3M, 24. Procter \& Gamble,
25. The Home Depot, 26. Mcdonald's, 27.  Alcoa, 28. Chevron Corporation,
29. United Technologies Corporation, 30. The Travelers Companies.

\end{document}